\documentstyle[12pt]{article}
\def\GA{\raise2.5pt\hbox{$>$}\kern-8pt\lower2.5pt\hbox{$\sim$}}
\def\LA{\raise2.5pt\hbox{$<$}\kern-8pt\lower2.5pt\hbox{$\sim$}}
\newcommand{\R}{{\cal R}}
\newcommand{\I}{{\cal I}}

\title{Exact Solutions of the One-Dimensional Quintic Complex Ginzburg-Landau
Equation}

\vspace{2.0cm}

\author{\\ \\Philippe Marcq, Hugues Chat\'e and Robert Conte\\ \\ \\ \\
\normalsize
Service de Physique de l'Etat Condens\'e\\
\normalsize
Centre d'Etudes de Saclay\\
\normalsize
91191 Gif-sur-Yvette, France.}
\date{\mbox{ }}

\begin{document}
\maketitle
\vskip 1 cm
\begin{abstract}
\noindent
Exact solitary wave solutions of the one-dimensional
quintic complex Ginzburg-Landau equation
are obtained using a method
derived from the Painlev\'e test for integrability.
These solutions are expressed in terms of hyperbolic
functions, and include the pulses and fronts
found by van Saarloos and Hohenberg. We also find previously
unknown sources and sinks.
The emphasis is put on the systematic character of the method which breaks
away from approaches involving somewhat ad hoc Ans\"atze.
\baselineskip 18pt
\end{abstract}

\newpage

\section{Introduction}

A number of non-integrable, non-linear, dissipative
partial differential equations (PDEs) are known to display
a wide variety of complex behavior, where the global time evolution is often
governed by the dynamics of spatially localized structures.
For example, in defect-mediated turbulence~\cite{lega}, the disordered
creation, motion and annihilation of topological defects play a prominent role.
In some cases of spatiotemporal intermittency, well-defined
localized objects ``carry''
the disorder and act as spatial delimiters of laminar regions~\cite{sti}.
This was shown  in particular
for the Nozaki-Bekki family of exact solutions of the
one-dimensional supercritical complex Ginzburg-Landau equation
(see below, Eq.~(\ref{cgl3}))~\cite{cglsti}.
More strongly disordered regimes of simple PDEs like the Kuramoto-Sivashinsky
equation, in which localized objects do not appear in an obvious
manner, have been argued to be well-described as a ``gas'' of interacting
pulses which are themselves exact solutions of the governing
equation~\cite{toh}.
\baselineskip 18pt

All these objects are, to a variable extent, reminiscent of the solitons
of completely integrable equations. They can be thought of (at least on
a qualitative level) as the dissipative
counterparts of invariant tori in chaotic Hamiltonian
systems, in the sense that they represent the preserved part of the
rich mathematical structure of nearby integrable systems.
Methods developed to investigate integrability of
differential equations
can therefore be expected to shed new light on spatiotemporally
chaotic dynamics \cite{seg}.

\vspace{24pt}
Following this line of thought, we present new exact
particular solutions for the quintic Complex Ginzburg-Landau
(CGL) equation in one spatial dimension, using techniques
derived from the Painlev\'e test for integrability \cite{ARS}.
The quintic CGL equation reads:
\begin{equation}
\label{cgl5}
{ \partial A \over \partial t} = \varepsilon A +
(b_1 + i c_1 ) {\partial^2 A \over \partial x^2}
- (b_3 - i c_3) |A|^2 A
- (b_5 - i c_5) |A|^4 A,
\end{equation}
where $\varepsilon$, $b_1$, $c_1$, $b_3$, $c_3$, $b_5$, $c_5$,
are real constants and the field $A(x ,t)$ is complex.

Eq.~(\ref{cgl5}) is a one-dimensional model of the large-scale behavior
of many nonequilibrium pattern-forming systems \cite{cross}.
When the real parameters $b_3$ and $b_5$ are respectively negative and
positive, Eq.~(\ref{cgl5}) accounts for the
slow modulations in space and time of an oscillatory mode close to a
subcritical
Hopf bifurcation. The discontinuous character of this symmetry-breaking
bifurcation is responsible for the occurrence of  metastable states
separated by fronts \cite{pom}.
Pulse solutions have also been argued to exist and play an important
dynamical role~\cite{pulse}. Examples of relevant experimental contexts
include binary fluid convection \cite{kol} and Taylor-Couette flow
between counter-rotating cylinders \cite{heg}.

We build on the work of W. van Saarloos and P.C. Hohenberg \cite{vsh},
who recently reviewed the properties of the solutions of Eq.~(\ref{cgl5})
which they called {\it coherent structures} in order to emphasize
their strong, usually exponential, spatial localization.
In order to attempt to stop the proliferation of notations,
we have chosen to use their notation throughout this paper, as well as their
vocabulary to distinguish among the various types of coherent structures.

Exact, analytical solutions of the CGL equation are scarce \cite{vsh,kly,pow},
and in any case limited to the uniformly propagating case,
{\it i.e.} solutions of the form:
\begin{equation}
A(x,t) = e^{-i \omega t} \ \hat{A}(\xi = x - v t).
\end{equation}
The original PDE, depending upon ($x$,\ $t$), is thus reduced to a second-order
ordinary differential equation (ODE) in the $\xi = x - v t$
independent variable, where $v$ is a constant velocity,
and ``$\;'\;$'' stands for differentiation with respect to $\xi$:
\begin{equation}
\label{ode}
- v {\hat A}' = (\varepsilon + i \omega) {\hat A} +
(b_1 + i c_1 ) {\hat A}''
- (b_3 - i c_3) |{\hat A}|^2 {\hat A}
- (b_5 - i c_5) |{\hat A}|^4 {\hat A}.
\end{equation}

Eq.~(\ref{ode}) possesses two types of fixed point:
linear fixed points which correspond to
a trivial vacuum state, $A(x,t) = 0$, and non-linear fixed points
which correspond to plane waves:
$$A(x,t) = a_N e^{-i \omega_N t + i q_N x}.$$

The next step consists in looking for
connections between any two of these elementary
objects in the phase space of Eq.~(\ref{ode}).
These connecting objects, the coherent structures
mentioned above, are classified as follows:
\begin{itemize}
\item[-]{pulses, {\it i.e.} homoclinic orbits between
two vacuum states (linear fixed points) at \hbox{$ \xi = \pm \infty$},}
\item[-]{fronts, {\it i.e.} heteroclinic orbits
between a vacuum state at $\pm \infty$ and a plane wave (non-linear fixed
point)
at $\mp \infty$,}
\item[-]{sources, {\it i.e.} heteroclinic orbits
between two outgoing waves at $\pm \infty$,}
\item[-]{sinks, {\it i.e.} heteroclinic orbits
between two incoming waves at $\pm \infty$.}
\end{itemize}
Sources and sinks can be distinguished from each other
by checking the signs of the group velocities of the asymptotic
plane waves at $\pm \infty$.
The multiplicity of these coherent structures admits an upper bound, determined
by
the dimensionality of the connecting manifolds, flowing from the
unstable eigendirection(s) of one fixed point into the stable
eigendirection(s) of another fixed point.

Exact solutions to Eq.~(\ref{ode}) were obtained in \cite{vsh} by using
an ad hoc, reduction of order Ansatz, where the first-order derivatives
of the phase and amplitude of $A(\xi)$ are given {\it a priori} expressions.
Van Saarloos and Hohenberg found exact pulses and fronts,
but did not mention the existence
of any exact source or sink for the quintic equation, although such solutions
are allowed by the counting arguments.
Finally, they  gave numerical evidence of the
important role played by these special, highly non-generic solutions,
which were shown to be ``dynamically selected'' in certain
regions of parameter space for sufficiently localized
initial conditions.

\vspace{24pt}

The purpose of this paper is to show that
a coherent mathematical framework can foster a better understanding
of these exact solutions and of the precise reasons
for their functional form.
The emphasis is laid upon a {\it local}
study of the {\it analytical} structure of possible solutions
in the complex plane, as opposed to the more geometrical methods
referred to above. Similar arguments were recently used by
Conte and  Musette in the case of the cubic CGL equation \cite{cgl3},
for which the quintic term vanishes:
\begin{equation}
\label{cgl3}
{ \partial A \over \partial t} = \varepsilon A +
(b_1 + i c_1 ) {\partial^2 A \over \partial x^2} - (b_3 - i c_3) |A|^2 A.
\end{equation}
All the known solutions of (\ref{cgl3}) were naturally retrieved in
\cite{cgl3},
including the one-parameter family of sources originally due to Nozaki
and Bekki \cite{nozbek}.

In addition to the known pulses and fronts of \cite{vsh},
we find for the quintic CGL equation
a new set of sinks and sources, whose existence is restricted to a
low co-dimension subspace of the full \hbox{($\varepsilon$, $b_1$, $c_1$,
$b_3$, $c_3$, $b_5$, $c_5$)} parameter space.
These three types of solution locally obey the same singularity structure,
and use hyperbolic functions as the elementary units from
which their global functional form is built.
However, their respective multiplicities do not reach the
upper bounds set in \cite{vsh}.

\section{Methodology}

\subsection{The Painlev\'e test for integrability}

Our guideline will be integrability in the sense of Painlev\'e:
an ODE will be called integrable if its general solution is free
from {\it movable critical} points.

Let us first define these adjectives. A critical point
(of a complex-valued application) is a point around which several
determinations of the application occur. Examples include algebraic and
logarithmic branch points.

A movable singular point (of a solution
of a DE) is a singular point whose location in the complex plane is not
determined by the coefficients of the DE. This location can only be
obtained from the initial conditions of the differential
problem, {\it i.e.} from integration constants. The simplest example
is provided by the Bernoulli equation:
$$u'(x) = - u(x)^2,$$
whose general solution
$$u(x) = {1 \over x - x_0}$$
admits a movable simple pole at $x_0$.
Conversely, singular points whose location depends only upon
the coefficients of the DE are called fixed.
Linear DE's only admit fixed critical points.

In this context, integrability is intimately connected with single-valuedness:
integrating a differential equation is ultimately equivalent to expressing its
solution in terms of functions, {\it i.e.} single-valued applications
of {\em \bf C} into {\em \bf C}. Multi-valuedness, expressed through the
occurrence of critical points, can be easily dispensed with when the
critical points are fixed, for instance by removing from the domain of
the solution a line in {\em \bf C} between the critical point and a point
at infinity. On the other hand, movable critical points are
sources of persistent multi-valuedness, and therefore preclude integrability.

Implicit in this scheme is
the necessity of extending the domain of independent variables to the
complex plane. Although unphysical at first sight, this prerequisite
is simply analogous to solving real algebraic equations for complex
unknowns.

The strength of the Painlev\'e test for an ODE
\begin{equation}
\label{eqex}
E\left[\;{\hbox{d} \over \hbox{d}x}, \ u(x)\right] = 0
\end{equation}
lies in its easy, algorithmic implementability. Its
main requirement is the existence of all possible solutions $u(x)$ of
Eq.~(\ref{eqex})
expressed as a Laurent expansion in a neighborhood of a
movable singularity $x_0$:
\begin{equation}
\label{lau}
u(x) = \chi(x)^{-\alpha} \ \sum_{j=0}^{\infty} u_j \ \chi(x)^j,
\end{equation}
where $\alpha$ is the leading-order exponent, $\chi(x)$
the expansion variable, and $\{ u_j, \;j \ge 0\}$ a set
of constant coefficients.
Following the invariant formulation of Conte \cite{inv}, we
distinguish here the expansion variable
$\chi(x)$ from the singular manifold of Weiss, Tabor, and Carnevale
\cite{ARS}, and only require $\chi(x)$ to behave as a
simple zero near $x_0$: $\chi(x) \sim x-x_0$.
This expansion is substituted into $E[\hbox{d}/\hbox{d}x,u(x)]$.
The $u_j$'s are determined from recursion relations
that develop when the coefficients at each order of $\chi$ are
required to vanish. The leading-order exponent $\alpha$ is determined
by equating the exponents of the dominant order terms in the
DE (\ref{eqex}).

\noindent
Necessary conditions for an ODE to pass the Painlev\'e test are:
\begin{itemize}
\item[1.]{the leading order $\alpha$ is an integer,}
\item[2.]{the recursion relation for the coefficients
$u_j$ can be consistently solved to any order,}
\end{itemize}
and possibly some other conditions not detailed here \cite{cfp}.
This procedure checks that the Laurent-type expansion
for $u(x)$ (Eq.~(\ref{lau})) is both consistent and
free from logarithmic branch points.

\subsection{Painlev\'e analysis for nonintegrable equations}

The general solution of nonintegrable equations will fail the
Painlev\'e test at one of these two steps. However, this does not forbid
the existence of particular solutions, provided that they respect
the singularity structure derived from the leading-order analysis.

The next step consists in determining how the local, analytical structure
valid in a neighborhood of a singular point $x_0$ can be taken into account
to yield global results, namely expressions of $u(x)$ valid
for {\em \bf C} as a whole. Once again, the guideline is provided by
one of the early results of Painlev\'e on integrability \cite{pain1}:
the solutions of all known integrable non-linear ODEs
of order at most two and degree one in $u''(x)$ can be expressed as linear
combinations
of logarithmic derivatives of entire functions, whose coefficients
are entire functions \cite{cours}. We mention here as an example
the case of (P2), one of the six integrable second-order equations
of Painlev\'e:
$$ u'' = 2 u^3 + x u + a$$
where $a$ is a constant coefficient. Its general solution can
be expressed as:
$$ u(x) = \partial_{x} \hbox{Log} \psi_1 - \partial_{x} \hbox{Log} \psi_2, $$
$\psi_1$ and $\psi_2$ being two entire functions.

This result should not come as a surprise,
since $\partial_{x} \hbox{Log} \psi$ is by construction
single-valued, and behaves like a pole.
In this respect, the integrable (thus single-valued) part of nonintegrable
equations
is naturally expected to be expressible in terms of logarithmic derivatives
of entire functions.

We now turn to the definition of the class of possible solutions we
consider. Arguments will remain mostly heuristic, although our
Ansatz can be derived within a rigorous mathematical
setting, taking into account the inherent invariance of
Painlev\'e analysis under the group of homographic transformations.
For more details, we refer the more mathematically-oriented reader to
the articles \cite{cgl3} and \cite{inv} and to the lecture
notes \cite{cours}.

\subsection{Ansatz for the quintic CGL equation}
\label{hyp}

Leading-order analysis for Eq.~(\ref{ode}) is achieved by balancing the
highest-order derivative with the strongest nonlinearity.
${\hat A}(\xi)$ being a complex field, this must be done by writing two
complex conjugate equations for ${\hat A}(\xi)$ and
${\hat B}(\xi) = {\hat A}^{*}(\xi)$, where ``$^{*}$'' denotes complex
conjugation. The fields ${\hat A}$ and ${\hat B}$ are now formally
considered as independent variables, and obey:
\begin{equation}
\label{an1}
\begin{array}{lll}
(b_1 + i c_1) {\hat A}'' &\sim& (b_5 - i c_5) {\hat A}^3 {\hat B}^2,\\
(b_1 - i c_1) {\hat B}'' &\sim& (b_5 + i c_5) {\hat A}^2 {\hat B}^3.
\end{array}
\end{equation}
Using $\chi(\xi) \sim \xi-\xi_0$ and $\chi'(\xi) \sim 1$, and feeding the
leading-order Ansatz ${\hat A} \sim A_0 \ \chi^{\alpha}$,
${\hat B} \sim B_0 \ \chi^{\beta}$ into Eqs. (\ref{an1}) leads to:
\begin{equation}
\label{sing}
\begin{array}{lll}
{\hat A} &\sim& A_0 \ \chi^{- {\frac{1}{2}} + i \alpha_0},\\
{\hat B} &\sim& B_0 \ \chi^{- {\frac{1}{2}} - i \alpha_0},
\end{array}
\end{equation}
where $A_0$, $B_0$ and $\alpha_0$ are solutions of the equations:
\begin{eqnarray*}
&\I \ \alpha_0^2 + 2 \R \ \alpha_0 - \frac{3}{4} \I = 0,\\
&(A_0 B_0)^2 = 2 \ {b_1^2 + c_1^2 \over \I} \ \alpha_0,
\end{eqnarray*}
and the intermediate variables $\R$ and $\I$ are defined as:
\begin{eqnarray*}
\R &= {\mbox Re}\ [\ (b_1 + i c_1) (b_5 + i c_5)\ ] &= b_1 b_5 - c_1 c_5,\\
\I &= {\mbox Im}\ [\ (b_1 + i c_1) (b_5 + i c_5)\ ] &= b_1 c_5 + c_1 b_5.
\end{eqnarray*}
Without loss of generality, we assume that $A_0 = B_0$ are real constants
in the rest of this paper.

We consider here the non-degenerate case $\I \neq 0$, where:
\begin{equation}
\label{a0}
\begin{array}{lll}
\alpha_0 &=& -{\frac{\R}{\I}} \pm \sqrt{ {\frac{3}{4}} + \left(
{\frac{\R}{\I}} \right) ^2}\;\neq 0,\\
A_0^4 &=& 2 \ {b_1^2 + c_1^2 \over \I} \ \alpha_0.
\end{array}
\end{equation}
The degenerate case $\I=0,\ \alpha_0 = 0$ is treated in the Appendix.
It should be noted that all results we present
(including the Appendix) respect the leading-order balance
(\ref{an1}), and are thus valid only when the coefficients of the
quintic term and of the highest order derivative are both
non-zero:
\begin{equation}
\label{nondeg}
\left\{
\begin{array}{lll}
b_1 + i c_1 &\neq& 0,\\
b_5 + i c_5 &\neq& 0.
\end{array}
\right.
\end{equation}

The leading-order exponent $-\frac{1}{2} + i \alpha_0$
is not an integer: the nonintegrable complex
\hbox{Ginzburg-Landau} equation has already failed
the Painlev\'e test. As expected, its general solution
cannot be expressed in terms of elementary functions.
However, partial integrability
remains possible, in so far as one looks for solutions exhibiting
minimal - but necessary - multi-valuedness,
including movable algebraic and logarithmic branch points.
This requirement is clearly fulfilled by the following
rewriting of Eq.~(\ref{sing}):
\begin{equation}
\label{a1}
A(\xi) = A_0 \ e^{-i \omega t} \
R(\xi)^{\frac{1}{2}} \ e^{i \alpha_0 \Theta(\xi)}.
\end{equation}
We introduced here a generalized amplitude $R(\xi)$ and a generalized phase
$\Theta(\xi)$, {\it a priori} complex-valued. The amplitude
$R(\xi)$ is assumed to have at worst
the singular behavior of a pole, $\Theta(\xi)$ that of a logarithmic
branch point.

The global structure of possible solutions is introduced as follows.
To a given leading-order exponent $\alpha_0$ correspond four distinct
values of $A_0$, related through a phase shift of $\frac{\pi}{2}$.
Four families of entire functions $\psi_i$, $i=1,...,4$
must therefore be introduced in the expression of $R(\xi)$:
\begin{equation}
\label{essai1}
R(\xi) = r_0(\xi) + \sum_{i=1}^4 r_i(\xi) \ \partial_{\xi} \hbox{Log} \psi_i.
\end{equation}
The coefficients $r_i$, $i=0,...,4$ are assumed to be entire functions of
$\xi$,
thus ensuring a pole behavior for $R(\xi)$ in the vicinity of any of the
movable zeroes of the $\psi_i$'s.
For simplicity, we will restrict our Ansatz to the case where:
\begin{itemize}
\item[-]{only two families $\psi_1$ and $\psi_2$ are used, and}
\item[-]{the coefficients $r_0$, $r_1$ and $r_2$ are real constants,}
\end{itemize}
thus keeping the number of unknowns at a tractable level with
respect to the number of equations.
Leading-order analysis leads to $r_1 = \pm r_2 = \pm 1$,
when conducted according to Eq.~(\ref{essai1})
with $r_0$, $r_1$, $r_2$ real constants, and $r_3 = r_4 = 0$.
Up to a constant phase shift, we can fix $r_1 = +1$, $r_2 = \pm 1$.
Our Ansatz for $R(\xi)$ reads:
\begin{equation}
\label{a2}
R(\xi) = r_0 + \partial_{\xi} \hbox{Log} \psi_1 \pm \partial_{\xi} \hbox{Log}
\psi_2.
\end{equation}
Let us now turn to the Ansatz for $\Theta(\xi)$.
Only one family is necessary here, which we denote $\psi_1$.
Since the complex Ginzburg-Landau equation is invariant under an
homogeneous phase translation $A \rightarrow A \ e^{i \phi}$,
$\Theta$ only contributes through its gradient $\partial_{\xi} \Theta$.
We only need to define an expression for $\partial_{\xi} \Theta$, with,
again, a pole singular behavior at worst. This expression reads:
\begin{equation}
\label{a3}
\partial_{\xi} \Theta(\xi) = \theta_0 + \partial_{\xi} \hbox{Log} \psi_1,
\end{equation}
where a constant coefficient $\theta_0$ was introduced.

The entire functions $\psi_1$ and $\psi_2$ are next defined
as solutions of integrable differential equations. For simplicity,
we assume that the $\psi_i$'s are solutions to the second-order
linear ODE:
\begin{equation}
\label{a4}
{\hbox{d}^2 \psi_{i} \over \hbox{d} \xi^2} = \frac{k^2}{4} \ \psi_i.
\end{equation}
Linear first-order differential equations have been excluded. Their
solutions have constant logarithmic derivatives, leading to
trivial expressions of $R(\xi)$ and $\Theta(\xi)$ which correspond
to the fixed-point solutions of Eqs. (\ref{ode}). Other
choices of the defining ODE are possible
in principle, but were at first discarded, again for reasons of simplicity.
This helped in keeping the algebraic
manipulations needed later at a tractable level.

The general solution to Eq.~(\ref{a4}) reads:
\begin{equation}
\label{sol}
\psi(\xi) = \psi_0 \ \cosh \frac{k}{2} (\xi - \xi_0),
\end{equation}
where $\psi_0$ and $\xi_0$ are the two integration constants.
The value of $\psi_0$ can be set to $1$, since only the ratio
$\partial_{\xi} \psi_i / \psi_i$ contributes. We define the two families as
two independent solutions to Eq.~(\ref{sol}) separated
by a constant phase shift denoted $k a$:
\begin{equation}
\label{phase}
\begin{array}{lll}
\psi_1(\xi) &=& \cosh \frac{k}{2} (\xi - \xi_0 + a),\\
\psi_2(\xi) &=& \cosh \frac{k}{2} (\xi - \xi_0 - a).
\end{array}
\end{equation}

Elementary manipulations then show that \cite{cgl3}:
\begin{eqnarray}
\label{th}
\partial_{\xi} \hbox{Log} \psi_1 \ +\
\partial_{\xi} \hbox{Log} \psi_2 &=&
{k \ \sinh k (\xi - \xi_0) \over \cosh k (\xi - \xi_0) \ + \
\cosh(k a)} ,\\
\label{sech}
\partial_{\xi} \hbox{Log} \psi_1 \ -\
\partial_{\xi} \hbox{Log} \psi_2 &=&
{k \ \sinh (k a) \over \cosh k (\xi - \xi_0) \ + \
\cosh(k a)}.
\end{eqnarray}
Our Ansatz (\ref{phase}) can be expected to yield
all possible solutions to the complex Ginzburg-Landau equation
involving hyperbolic functions or, up to linear combinations,
exponentials. Hyperbolic tangents can be obtained from \hbox{Eq.~(\ref{th})}
when $\cosh(k a) = 0$, {\it i.e.} $ka = i \frac{\pi}{2}$, hyperbolic secants
from \hbox{Eq.~(\ref{sech})} when $\cosh(k a) = 0$.

Eq.~(\ref{th}) shows that two families having identical coefficients
are in practice equivalent to one family, up to dividing the wavenumber $k$ by
$2$
and to setting $ka = i \frac{\pi}{2}$. This allows a
slight modification of the equation defining $\Theta(\xi)$,
formally using the two families $\psi_1$ and $\psi_2$.
The complete Ansatz for $A(\xi)$ now reads:
\begin{equation}
\label{ansatz}
\left\{
\begin{array}{lll}
A(\xi) &=& A_0 \ e^{-i \omega t} \
R(\xi)^{\frac{1}{2}} \ e^{i \alpha_0 \Theta(\xi)},\\
R &=& r_0 \ +\ \partial_{\xi} \hbox{Log} \psi_1 \ \pm \
\partial_{\xi} \hbox{Log} \psi_2,\\
\partial_{\xi} \Theta &=& \theta_0 \ + \ \partial_{\xi} \hbox{Log} \psi_1 \ +
\ \ \partial_{\xi} \hbox{Log} \psi_2,\\
\partial_{\xi \xi} \psi_i &=& \frac{k^2}{4} \ \psi_i, \ \ i=1,2.
\end{array}
\right.
\end{equation}
Albeit very restrictive, this Ansatz suffices to retrieve all the
exact solutions quoted in \cite{vsh}.

\subsection{Computational aspects}
\label{comp}

{}From now on, the constant coefficients $\theta_0$, $\omega$, $c$
will be real numbers, in order to avoid unbounded solutions.
The wavenumber $k$ is also assumed to be real.
Possible periodic solutions are thus
discarded, since they lack the required spatial localization.
On the other hand, taking $r_0 \in$ {\em \bf C} would not restrict the
generality
of Ansatz (\ref{ansatz}). However,we checked that the quintic equation,
unlike the cubic one (see \cite{cgl3}), does not admit any
solutions respecting the less restrictive hypothesis $r_0 = x_0 + i\; y_0$;
$x_0$, $y_0 \in$ {\em \bf R}; $x_0 y_0 \neq 0$.

The basic principle of our resolution is to turn the differential
equation we want to solve into a much simpler, purely algebraic problem.
This is made possible by the analytic considerations of the previous
section: the spatial structure of solutions to Eq.~(\ref{ode})
is supposed to be fully contained in the elementary logarithmic
derivatives $\partial_{\xi} \hbox{Log} \psi_i$, whose
functional form will never be made explicit in our computations. The
sometimes intricate algebraic manipulations can then be solved quite easily
with the help of any symbolic mathematics package, such as
Mathematica \cite{math}, or AMP \cite{amp}.

We first substitute our general Ansatz (Eqs.~(\ref{ansatz})) into
Eq.~(\ref{ode}) and successively eliminate all derivatives
of $\psi_i$ of order greater than or equal to two by using:
$$
{\hbox{d}^2 \psi_{i} \over \hbox{d} \xi^2} = \frac{k^2}{4} \ \psi_i.
$$
Eq.~(\ref{ode}) is then equivalent to a polynomial equation in the
$\partial_{\xi} \hbox{Log} \psi_1$ and
$\partial_{\xi} \hbox{Log} \psi_2$ variables:
\begin{equation}
\label{cross}
\sum_{k=0}^{4} \ \sum_{m+n = k} F_{k} \
\left( \partial_{\xi} \hbox{Log} \psi_1 \right)^m \
\left( \partial_{\xi} \hbox{Log} \psi_2 \right)^n = 0,
\end{equation}
where the coefficients $F_k$ depend algebraically on
the parameters ($\varepsilon$, $b_i$, $c_i$),
on the unknowns $\omega$, $k$, $v$, $a$, $r_0$, $\theta_0$, and on
$A_0$ and $\alpha_0$, whose values are known from Eq.~(\ref{a0}).

A convenient way of taking the phase shift $k a$ into account
is to use a new variable $\mu_0$,
defined as \hbox{$\mu_0 = \coth(k a)$}. The products
$(\partial_{\xi} \hbox{Log} \psi_1)^m \ (\partial_{\xi} \hbox{Log} \psi_2)^n$
can be recursively linearized, from $m+n = 4$ to $m+n = 2$, by means of
the following identity:
\begin{equation}
\label{mu0}
\partial_{\xi} \hbox{Log} \psi_1 \ \partial_{\xi} \hbox{Log} \psi_2 =
\frac{k^2}{4} - \; \mu_0 \; \frac{k}{2} \
\left[ \;
\partial_{\xi} \hbox{Log} \psi_1 - \partial_{\xi} \hbox{Log} \psi_2\;
\right].
\end{equation}
The coefficient $\mu_0$ is now a real constant,
in agreement with the previous assumptions. We obtain:
\begin{equation}
\label{pol}
\sum_{j=1}^{4} E_{-j} \
\left( \partial_{\xi} \hbox{Log} \psi_1 \right)^j \ + \ E_0 \ + \
\sum_{j=1}^{4} E_{j} \
\left( \partial_{\xi} \hbox{Log} \psi_2 \right)^j \ = 0.
\end{equation}

Solving Eq.~(\ref{pol}) amounts to canceling all the $E_j$ coefficients,
$j = -4, \ldots, 4$. This is done recursively,
from $E_{-4}$ and $E_4$ to $E_0$, by a triangularization
technique. Parameters and unknowns are considered on
an equal footing, as variables whose values are successively
obtained from $E_j$ and substituted into equations $E_i$,
$0 \leq |i| < |j|$, thus decreasing the number of unknowns
in equations of lower index. For each $j$, the pivoting variable was
determined by singling out which variable admits the simplest expression
with respect to all other variables, while excluding possibly
vanishing denominators. These selected variables are:
\begin{itemize}
\item[-]{$a_5 = b_5 - i c_5$, obtained from $E_{-4}$ or $E_{4}$,}
\item[-]{$a_3 = b_3 - i c_3$, obtained from $E_{-3}$,}
\item[-]{${\hat v} = v + 2 i \alpha_0 a_1 \theta_0$, obtained from $E_{3}$,
where $a_1 = b_1 + i c_1$, and}
\item[-]{${\hat \varepsilon} = \varepsilon + i \theta_0 \omega$,
obtained from $E_{-2}$.}
\end{itemize}
Equation $E_2$ is then seen to vanish, and
the remaining $E_{-1}$, $E_0$ and $E_1$ equations depend only on the
parameters $b_1$, $c_1$, the leading-order quantities $A_0$, $\alpha_0$,
and the unknowns $k$, $\mu_0$ and $r_0$. Systematic resolution of these
equations
lead to the three cases detailed below,
where the explicit values of the real unknowns
$k$, $v$, $\omega$, $\theta_0$, $r_0$ and $\mu_0$ are given as functions
of the parameters \hbox{($\varepsilon$, $b_i$, $c_i$)},
$A_0$ and $\alpha_0$. The leading-order quantities
$A_0$ and $\alpha_0$ are considered as parameters, since their value can be
expressed
as functions of $b_1$, $c_1$, $b_5$, and $c_5$ (Eq.~(\ref{a0})).

\section{Results}
\label{res}

We do not mention here ``unphysical'' solutions obtained
for complex values of the parameters $k$, $v$, $\omega$, $\theta_0$, $r_0$ or
$\mu_0$, although such solutions may well become ``physical'' when more general
hypotheses are considered. These questions have been left for future work.

\subsection{Pulses}

The pulse solutions given in \cite{vsh} can be obtained from the following
Ansatz:
$$
\left\{
\begin{array}{lll}
R &=& \partial_{\xi} \hbox{Log} \psi_1 \ - \
\partial_{\xi} \hbox{Log} \psi_2,\\
\partial_{\xi} \Theta &=& \theta_0 \ + \ \partial_{\xi} \hbox{Log} \psi_1
\ + \ \partial_{\xi} \hbox{Log} \psi_2,
\end{array}
\right.
$$
where $r_0$ was set to $0$ in order to respect the
suitable asymptotic behavior:
$$
\lim_{\xi \rightarrow - \infty} A(\xi) =
\lim_{\xi \rightarrow + \infty} A(\xi) = 0.
$$
The solution reads:
\begin{eqnarray}
\label{pulse}
A(x,t) &=& A_0 e^{- i \omega t} e^{i \alpha_0 \theta_0 \xi}
\left[ \ \cosh k(\xi - \xi_0) + \cosh ka \ \right]^{i \alpha_0}\\
\nonumber
&&\times \left[ {k \ \sinh (k a) \over \cosh k (\xi - \xi_0) \ + \ \cosh(k a)}
\right]^{\frac{1}{2}}.
\end{eqnarray}
In this particular case, triangularizing the equations $E_j$, $j = -4,...,4$
leads to ${\hat v} = 0$, or
\hbox{$v = - 2 i \alpha_0 (b_1 + i c_1) \theta_0$}.
The velocity $v$ being a real constant, we obtain:
\begin{eqnarray*}
b_1 \theta_0 &=& 0,\\
v &=& 2 c_1 \alpha_0 \theta_0.
\end{eqnarray*}
We find a set of solutions, discrete when $b_1 \neq 0, \; v = \theta_0 = 0$,
parametrized by the velocity $v$ or by $\theta_0$ when $b_1 = 0$,
in a co-dimension-one subspace of parameter-space defined by:
$$
c_3 \left[ b_1 (1 - 2 \alpha_0^2) + 3 \alpha_0 c_1 \right] =
b_3 \left[ 3 \alpha_0 b_1 + c_1 ( 2 \alpha_0^2 - 1) \right].
$$
The unknowns $k$, $\mu_0$ and $\omega$ can be computed from:
\begin{eqnarray*}
k^2 &=& - {4 \varepsilon \over b_1 (1 - 4 \alpha_0^2) + 4 \alpha_0 c_1},\\
k \mu_0 &=& {- b_3 A_0^2 \over
b_1 (1 - 2 \alpha_0^2) + 3 \alpha_0 c_1 },\\
\omega &=& - c_1 \alpha_0 \theta_0^2 +
{\varepsilon \over \alpha_0} \; {-4 \alpha_0 b_1
+ c_1 (1- 4 \alpha_0^2) \over b_1 (1 - 4 \alpha_0^2) + 4 \alpha_0 c_1}.
\end{eqnarray*}

\subsection{Fronts}

The front solutions of \cite{vsh} involve hyperbolic tangents.
The Ansatz we use goes as follows:
$$
\left\{
\begin{array}{lll}
R &=& r_0 \ +\ \partial_{\xi} \hbox{Log} \psi_1 \ + \
\partial_{\xi} \hbox{Log} \psi_2,\\
\partial_{\xi} \Theta &=& \theta_0 \ +\ \partial_{\xi} \hbox{Log} \psi_1
\ + \ \partial_{\xi} \hbox{Log} \psi_2.\\
\end{array}
\right.
$$
Solutions were found only when $\mu_0 = 0$. Their explicit form reads:
\begin{equation}
\label{front}
A(x,t) = A_0 e^{- i \omega t} e^{i \alpha_0 \theta_0 \xi}
\left[\cosh k(\xi - \xi_0)\right]^{i \alpha_0}
\left[ k ( \tanh k(\xi - \xi_0) \ \pm \ 1 )\right]^{\frac{1}{2}},
\end{equation}
where the appropriate asymptotic behavior is obtained by setting $r_0$
to $\pm k$:
\begin{eqnarray*}
\lim_{\xi \rightarrow \mp \infty} A(\xi) &=& 0,\\
\lim_{\xi \rightarrow \pm \infty} A(\xi) &=& \sqrt{\pm 2 k} A_0
e^{- i \omega t} e^{i \alpha_0 (\theta_0 \pm k)}.
\end{eqnarray*}
Its variables obey different relationships according to
the value of $b_1$:
\begin{itemize}
\item[-]{
{\bf First subcase}: $b_1 \neq 0$. A discrete set of solutions is found.
The wavenumber $k$ is determined by a quadratic equation:
$$
a_k k^2 + b_k k + c_k = 0,
$$
where:
$$
\left\{
\begin{array}{lll}
a_k &=& -(1 + 4 \alpha_0^2 )^2  \ (3 b_1^2  + 4 c_1^2 ),\\
b_k &=& \mp 2 A_0^2 \ (1 + 4 \alpha_0^2 ) \ (b_1 b_3 - 2 c_1 c_3 + 2 \alpha_0
(b_1 c_3 + 2 b_3 c_1)),\\
c_k &=& b_1 \varepsilon \;(1 + 4 \alpha_0^2)^2 - A_0^4 \;
(c_3^2 + 4 b_3^2 \alpha_0^2 -4 \alpha_0 b_3 c_3).
\end{array}
\right.
$$
The unknowns $\theta_0$, $v$ and $\omega$ are obtained as functions of $k$:
\begin{eqnarray*}
\theta_0 &=&
\pm k \; (1 - 2 {c_1 \over \alpha_0 b_1}) + A_0^2\;
{c_3 - 2 \alpha_0 b_3 \over \alpha_0 (1 + 4 \alpha_0^2) b_1},\\
v &=&
\mp\;2 {b_1^2 + c_1^2 \over b_1} k - 2 \;{A_0^2 \over b_1}\;
{(c_1 c_3 - b_1 b_3) + 2 \alpha_0 (b_1 c_3 + b_3 c_1) \over
1 + 4 \alpha_0^2},\\
\omega &=& {1 \over b_1^2 \alpha_0 (1 + 4 \alpha_0^2)^2}\;
[\;a_{\omega} k^2 + b_{\omega} k + c_{\omega}\;],
\end{eqnarray*}
where:
$$
\left\{
\begin{array}{lll}
a_{\omega} &=& -c_1 (1 + 4 \alpha_0^2 )^2 \; (5 b_1^2  + 4 c_1^2 ),\\
b_{\omega} &=& \pm \; 2 A_0^2 (1 + 4 \alpha_0^2 )\\
&&\times [ (b_1^2  c_3 + 2 c_1^2  c_3 - 2 b_1 b_3 c_1) - 2 \alpha_0 \;
(b_1^2 b_3  +2 b_3 c_1^2  + 2 b_1 c_1 c_3 )],\\
c_{\omega} &=& A_0^4\; (c_3 -2 \alpha_0 b_3) \;
[(2 b_1 b_3 -c_1 c_3) + 2 \alpha_0 (b_3 c_1 + 2 b_1  c_3 )].
\end{array}
\right.
$$
}
\item[-]{
{\bf Second subcase}: $b_1 = 0$. In the non-degenerate case treated here
($\I = b_1 c_5 + c_1 b_5 \neq 0$),
this condition implies that $c_1$ cannot vanish.
We find a one-parameter family of
solutions restricted to a co-dimension-two subspace of parameter space
defined by:
\begin{eqnarray*}
b_1 &=& 0,\\
c_1 (1 + 4 \alpha_0^2)^2 \; \varepsilon &=& (c_3 - 2 \alpha_0 b_3)
(b_3 + 2 \alpha_0 c_3) \; A_0^4.
\end{eqnarray*}
The unknowns $\theta_0$, $\omega$ and $k$ are parametrized by the velocity $v$:
\begin{eqnarray*}
\theta_0 &=& {v \over 2 \alpha_0 c_1}\;+\;
{A_0^2 \over 2 \alpha_0 c_1}\;{2 b_3 (1 - \alpha_o^2) + 5 \alpha_0 c_3
\over 1 + 4 \alpha_0^2},\\
\omega &=& -{v^2 \over 4 \alpha_0 c_1}\\
&&+ {A_0^4 \over 4 \alpha_0 (1 + 4 \alpha_0^2)^2 c_1} \;
[ (2 b_3 + 4 \alpha_0 c_3)^2 - (2 \alpha_0 b_3 - c_3)^2 ],\\
k &=& \pm \; {A_0^2 \over 2 c_1} \;
{c_3 - 2 \alpha_0 b_3 \over 1 + 4 \alpha_0^2}.
\end{eqnarray*}
}
\end{itemize}

\subsection{Sources and sinks}

Sources and sinks are obtained from the Ansatz:
$$
\left\{
\begin{array}{lll}
R &=& r_0 \ +\
\partial_{\xi} \hbox{Log} \psi_1 \ - \ \partial_{\xi} \hbox{Log} \psi_2,\\
\partial_{\xi} \Theta &=& \theta_0 \ +\
\partial_{\xi} \hbox{Log} \psi_1 \ + \ \partial_{\xi} \hbox{Log} \psi_2,\\
r_0 &\neq& 0.
\end{array}
\right.
$$
Solutions exist for $\mu_0 \neq 0$ only, and interpolate
between two plane waves:
\begin{eqnarray*}
\lim_{\xi \rightarrow - \infty} A(\xi) &=&  A_0 r_0 e^{-i \omega t}
e^{i \left[\  \alpha_0 (\theta_0 - k) \xi \ \right]},\\
\lim_{\xi \rightarrow + \infty} A(\xi) &=& A_0 r_0 e^{-i \omega t}
e^{i \left[\  \alpha_0 (\theta_0 + k) \xi \ \right]}.
\end{eqnarray*}
Their expression reads:
\begin{eqnarray}
\label{source}
A(x,t) &=& A_0 e^{- i \omega t} e^{i \alpha_0 \theta_0 \xi}
\left[\ \cosh k(\xi - \xi_0) + \cosh ka \ \right]^{i \alpha_0}\\
\nonumber
&&\times \left[ {k \ \sinh (k a) \over \cosh k (\xi - \xi_0) \ + \ \cosh(k a)}
+ r_0
\right]^{\frac{1}{2}}.
\end{eqnarray}
As for pulses, we obtain ${\hat v} = 0$, whence:
\begin{eqnarray*}
b_1 \theta_0 &=& 0,\\
v &=& 2 c_1 \alpha_0 \theta_0.
\end{eqnarray*}
We find a set of solutions, discrete when $b_1 \neq 0, \; v = \theta_0 = 0$,
parametrized by the velocity $v$ when $b_1 = 0$, in a co-dimension-one subspace
of parameter-space defined by:
$$
\varepsilon = \frac{1}{4} \left[ b_1 (2 k^2 - 3 r_0^2) + 6 \alpha_0 c_1
(k^2 - r_0^2) \right],
$$
where:
\begin{eqnarray*}
k^2 &=& {A_0^4 \over  \alpha_0^2 (1 + 4 \alpha_0^2)^2 (b_1^2+c_1^2)^2}
\left[\ 7 \alpha_0 (b_1 b_3 - c_1 c_3) +  (2 \alpha_0^2 - 3)
(b_3 c_1 + b_1 c_3) \ \right] \\
&&\times \left[\ 3 \alpha_0 (b_1 b_3 - c_1 c_3) +  (2 \alpha_0^2 - 1)
(b_3 c_1 + b_1 c_3) \ \right],\\
r_0 &=& {A_0^2 \over  \alpha_0 (1 + 4 \alpha_0^2) (b_1^2+c_1^2)}
\left[\ -3 \alpha_0 (b_1 b_3 - c_1 c_3) +  (1 - 2 \alpha_0^2)
(b_3 c_1 + b_1 c_3) \ \right].\\
\end{eqnarray*}
The remaining unknowns are given by:
\begin{eqnarray*}
\omega &=& - c_1 \alpha_0 \theta_0^2 +
\frac{1}{4} \left[ c_1 (2 k^2 - 3 r_0^2) - 6 \alpha_0 b_1 (k^2 - r_0^2)
\right],\\
\mu_0 &=& - {k^2 + r_0^2 \over 2 k r_0}.
\end{eqnarray*}
The asymptotic group velocities $v_{g,\; \pm}$ in a co-moving frame at
\hbox{$\xi \rightarrow \pm \infty$} are given by:
$$
v_{g,\; \pm}= {\partial \Omega \over \partial K_{\pm}},
$$
where the respective asymptotic pulsation and wavenumbers are $\Omega = \omega$
and \hbox{$K_{\pm} = \alpha_0 \;(\theta_0 \pm k)$}. We obtain:
\begin{eqnarray*}
v_{g,\; \pm} &=& {1 \over 2 \alpha_0} \left( \;
{\partial \omega \over \partial \theta_0} \pm
{\partial \omega \over \partial k} \; \right)\\
&=& - {v \over 2 \alpha_0} \pm {c_1 - 3 \alpha_0 b_1 \over 2 \alpha_0} k.
\end{eqnarray*}
In the generic case $b_1 \neq 0, \; v = \theta_0 = 0$, we find
$$
v_{g,\; +} = - v_{g,\; -} =  {c_1 - 3 \alpha_0 b_1 \over 2 \alpha_0} k,
$$
whose sign can be either positive or negative. These solutions can be either
sources or sinks.

\section{Conclusion}

The search for the functional form of the coherent structures
appearing in disordered regimes of extended,
non-equilibrium systems can be made systematic by
following two basic principles:
\begin{itemize}
\item[-]{the singularity structure of possible solutions of
the relevant DE's must be taken into account at an early stage,}
\item[-]{logarithmic derivatives of entire functions are the
elementary units from which exact solutions can be built.}
\end{itemize}

In this work, we followed these principles and derived systematically
an Ansatz for coherent structure solutions of the one-dimensional
quintic complex Ginzburg-Landau equation.
In addition to the already known pulses (\hbox{Eq.~(\ref{pulse}))} and
fronts \hbox{(Eq.~(\ref{front}))},
we found new source and sink solutions \hbox{(Eq.~(\ref{source}))}.
The corresponding computations involved only
the simplest possible Ansatz compatible with our framework.
We reduced the number of families of entire functions to two, used
real coefficients throughout, and chose to define the elementary entire
functions from the simplest available ODE. Within these restrictions,
exponentials appeared as the simplest building blocks
from which exact solutions can be formed. Relaxing these
constraints can be expected to yield new - and more complex -
solutions, provided that the resulting algebraic computations
are not made intractable by increasing the degree of equations to be solved.
These less restrictive Ans\"atze are currently under investigation.

The stability and dynamical relevance of the solutions, especially
the new source/sink, remain to be investigated. In \cite{pow},
a survey of the special analytical, topological and dynamical
properties of the highly non-generic fronts (Eq.~(\ref{front}))
was given, and the question of relating these three aspects
was raised. These properties are crucial steps in trying to understand the
spatiotemporally disordered regimes exhibited by the equation,
and deserve closer scrutiny. Another point of interest is the question
of a possible role played by these solutions
in regions of parameter space out of their domain
of existence. There, objects related to these solutions, but out of reach of
the simple
Ansatz used in this work, may appear as the relevant building blocks
in the (chaotic) dynamics.

To conclude, let us stress again the systematic character of the approach
taken and the potential extensions of the case treated here, not only
to less restrictive Ans\"atze, but also to other non-linear PDEs of
physical interest.

\vskip 2 cm
\section*{Appendix}

We treat here the degenerate case: ${\cal I} = b_1 c_5 + b_5 c_1 = 0$.
Leading-order analysis then leads to:
\begin{eqnarray}
\label{singdeg}
\alpha_0 &=& 0,\\
\nonumber
A_0^4 &=& \frac{3}{4} \; {b_1 b_5 - c_1 c_5 \over b_5^2 +
c_5^2},
\end{eqnarray}
whenever the condition (\ref{nondeg}) applies.
This singular behavior is taken into account by writing:
$$
A(\xi) = A_0 e^{-i \omega t} \ R(\xi)^{\frac{1}{2}} \ e^{i \Theta(\xi)},
$$
where $R(\xi)$ has at most the singular behavior of a pole,
and $\Theta(\xi)$ is a regular function. In the spirit of
Section \ref{hyp}, we write:
\begin{equation}
\label{ansdeg}
\left\{
\begin{array}{lll}
R &=& r_0 \ +\ \partial_{\xi} (\hbox{Log} \psi_1) \ \pm \
\partial_{\xi} (\hbox{Log} \psi_2),\\
\partial_{\xi} \Theta &=& \theta_0,\\
\partial_{\xi \xi} \psi_i &=& \frac{k^2}{4} \ \psi_i, \ \ i=1,2.
\end{array}
\right.
\end{equation}
As before, looking for particular solutions of the quintic CGL equation
with the restriction $b_1 c_5 + b_5 c_1 = 0$ amounts to solving
a system of algebraic equations. Its systematic resolution leads to
pulse, front, source and sink solutions. The method is identical to that
presented in Section \ref{comp}, and solutions are given within the
restrictions
of Section \ref{res}.

\subsection*{Pulses}

The pulse solutions read:
\begin{equation}
\label{pulsedeg}
A(x,t) = A_0 e^{i [ \theta_0 \xi - \omega t ]}
\left[ {k \ \sinh (k a) \over \cosh k (\xi - \xi_0) \ + \ \cosh(k a)}
\right]^{\frac{1}{2}}
\end{equation}
and necessarily respect ${\hat v} = 0$, $b_1 \theta_0 = 0$. We distinguish
the two cases:
\begin{itemize}
\item[-]{{\bf First subcase:} $b_1 \neq 0$. A discrete set of stationary pulses
is found in a co-dimension-two subspace of parameter space defined by:
$$
\begin{array}{lll}
b_1 c_5 + b_5 c_1 &=& 0,\\
b_1 c_3 + b_3 c_1 &=& 0.
\end{array}
$$
Such restrictions include the case of the Real Ginzbug-Landau equation
(RGL: $c_1 = c_3 = c_5 = 0$):
\begin{equation}
\label{rgl}
{\partial A \over \partial t} = \varepsilon A + b_1 {\partial^2 A \over
\partial x^2}
- b_3 |A|^2 A - b_5 |A|^4 A.
\end{equation}
All parameters are fixed:
$$
\begin{array}{lll}
v &=& 0,\\
\theta_0 &=& 0,\\
b_1 k^2 &=& - 4 \varepsilon,\\
b_1 k \mu_0 &=& - b_3 A_0^2,\\
b_1 \omega &=& \varepsilon c_1.
\end{array}
$$
}
\item[-]{{\bf Second subcase:} $b_1 = 0$. A two-parameter family of
pulses is found for the Quintic-Cubic Schr\"odinger
equation ($c_1 \neq 0$, $c_5 \neq 0$):
\begin{equation}
\label{nls}
- i {\partial A \over \partial t} = c_1 {\partial^2 A \over \partial x^2}
+ c_3 |A|^2 A + c_5 |A|^4 A.
\end{equation}
The free parameters are chosen to be the wavenumber $k$ and velocity $v$:
$$
\begin{array}{lll}
2 c_1 \theta_0 &=& v,\\
4 c_1 \omega &=& - v^2 - c_1^2 k^2,\\
c_1 k \mu_0 &=& A_0^2 c_3.
\end{array}
$$
}
\end{itemize}

\subsection*{Fronts}
Fronts are obtained for $\mu_0 = 0$ and $r_0 = \pm k$:
\begin{equation}
\label{frontdeg}
A(x,t) = k^{\frac{1}{2}} A_0 e^{i [ \theta_0 \xi - \omega t ]}
\left[ \ \tanh k(\xi - \xi_0) \ \pm \ 1 \ \right]^{\frac{1}{2}}.
\end{equation}
The condition $b_1 + i c_1 \neq 0$ leads us to distinguish two cases:
\begin{itemize}
\item[-]{{\bf First subcase:} $b_1 \neq 0$. A discrete set of fronts
is found in a co-dimension-one subspace of parameter space defined
by:
$$
b_1 c_5 + b_5 c_1 = 0.
$$
The wavenumber $k$ is a solution of the quadratic equation:
$$
a_k k^2 + b_k k + c_k = 0,
$$
where:
$$
\left\{
\begin{array}{lll}
a_k &=& (3 b_1^2 + 4 c_1^2),\\
b_k &=& \pm 2 A_0^2 \ (b_1 b_3 - 2 c_1 c_3),\\
c_k &=& - \varepsilon b_1 + c_3^2 A_0^4.
\end{array}
\right.
$$
The unknowns $\theta_0$, $v$ and $\omega$ are obtained as functions of $k$:
$$
\begin{array}{lll}
b_1 \theta_0 &=& \mp 2 c_1 k \; + A_0^2 c_3,\\
b_1 v &=& \mp 2 (b_1^2 + c_1^2) k + 2 A_0^2 (c_1 c_3 - b_1 b_3),\\
b_1^2 \omega &=& a_{\omega} k^2 + b_{\omega} k + c_{\omega},
\end{array}$$
where:
$$
\left\{
\begin{array}{lll}
a_{\omega} &=& -c_1 (5 b_1^2  + 4 c_1^2 ),\\
b_{\omega} &=& \pm \; 2 A_0^2 (b_1^2  c_3 + 2 c_1^2  c_3 - 2 b_1 b_3 c_1),\\
c_{\omega} &=& A_0^4\; c_3 \; (2 b_1 b_3 - c_1 c_3).
\end{array}
\right.
$$
}
\item[-]{{\bf Second subcase:} $b_1 = 0$, $c_1 \neq 0$.
A one-parameter family of solutions is found in a co-dimension-three
subspace defined by:
$$
\begin{array}{lll}
b_1 &=& 0,\\
b_5 &=& 0,\\
\varepsilon c_1 &=& A_0^4 b_3 c_3,
\end{array}
$$
thus including the generalized quintic-cubic NLS equation (Eq.~(\ref{nls}))
when $\varepsilon = b_3 = 0$.
All coefficients can be expressed as functions of the velocity $v$:
$$
\begin{array}{lll}
2 c_1 \theta_0 &=& v + 2 A_0^2 b_3,\\
4 c_1 \omega &=& -v^2 + A_0^4 (4 b_3^2 - c_3),\\
2 c_1 k &=& \pm A_0^2 c_3.
\end{array}
$$
}
\end{itemize}

\subsection*{Sources and sinks}

These solutions read:
\begin{equation}
\label{sourcedeg}
A(x,t) = A_0 e^{i [ \theta_0 \xi - \omega t ]}
\left[ r_0 + {k \ \sinh (k a) \over \cosh k (\xi - \xi_0) \ + \ \cosh(k a)}
\right]^{\frac{1}{2}},
\end{equation}
and respect ${\hat v} = 0$. We distinguish the two cases:
\begin{itemize}
\item[-]{
{\bf First subcase:} $b_1 \neq 0$. A discrete set of
stationary solutions
is found in a co-dimension-two subspace of parameter space defined by:
$$
\begin{array}{lll}
b_1 c_5 + b_5 c_1 &=& 0,\\
b_1 c_3 + b_3 c_1 &=& 0,
\end{array}
$$
including the RGL equation (Eq.~(\ref{rgl})).
All parameters are fixed: $r_0$ is determined by the quadratic equation:
$$
3 b_1^2 r_0^2 + 4 A_0^2 b_3 r_0 - 4 \varepsilon = 0,
$$
and:
$$
\begin{array}{lll}
v &=& 0,\\
\theta_0 &=& 0,\\
b_1 k^2 &=& - 2 A_0^2 b_3 r_0 + 4 \varepsilon,\\
b_1 \omega &=& \varepsilon c_1,\\
\mu_0 &=& - {k^2 + r_0^2 \over 2 k r_0}.
\end{array}
$$
The far-field group velocity $v_g$ vanishes, due to an asymptotic
stationary wave behavior:
$$
\lim_{\xi \rightarrow \pm \infty} A(x,t) = A_0 \sqrt{r_0}
e^{-i \omega t}.
$$
}
\item[-]{{\bf Second subcase:} $b_1 = 0$. A two-parameter family of
sources or sinks is found for the quintic-cubic NLS
equation (\ref{nls}).
We choose to use $v$ and $r_0$ as free parameters ($c_1 \neq 0$):
$$
\begin{array}{lll}
2 c_1 \theta_0 &=& v,\\
4 c_1 \omega &=& - v^2 + 3 c_1^2 r_0^2 - 4 A_0^2 c_1 c_3 r_0,\\
c_1 k^2 &=& 3 c_1 r_0^2 - 2 A_0^2 c_3 r_0,\\
\mu_0 &=& - {k^2 + r_0^2 \over 2 k r_0}.
\end{array}
$$
The asymptotic group velocities $v_{g,\;\pm}$ read:
$$
v_{g,\; +} = - v_{g,\; -} = {\partial \omega \over \partial \theta_0} = -v.
$$
}

\end{itemize}

\end{document}